\documentstyle[12pt]{article}

\def \la {{\langle}}
\def \ra {{\rangle}}

\newcommand{\be}{\begin{equation}}
\newcommand{\ee}{\end{equation}}
\newcommand{\beqar}{\begin{eqnarray}}
\newcommand{\eeqar}{\end{eqnarray}}

\newcommand{\Tr}{{\rm Tr}}
\newcommand{\half}{{\frac{1}{2}}}
\newcommand{\tr}{{\rm tr}}

\expandafter\ifx\csname mathbbm\endcsname\relax

\else

\fi
\begin{document}
\begin{titlepage}
\begin{flushleft}  
       \hfill                      CCNY-HEP-01-12\\
       \hfill                      RU-01-8-B\\
       \hfill                      {\tt hep-th/0106011}\\
       \hfill                      May 2001\\
\end{flushleft}
\vspace*{3mm}
\begin{center}
{\LARGE Quantum Hall states on the cylinder as unitary matrix 
Chern-Simons theory
\\}
\vspace*{12mm}
{\large Alexios P. Polychronakos\footnote{On leave from Theoretical 
Physics Dept., Uppsala University, 751 08 Sweden and Dept.~of Physics, 
University of Ioannina, 45110 Greece; E-mail: poly@teorfys.uu.se} \\
\vspace*{5mm}
{\em Physics Department, City College of the CUNY \\
New York, NY 10031, USA}\\
\vspace*{5mm}
and \\
\vspace*{5mm}
{\em Physics Department, The Rockefeller University \\
New York, NY 10021, USA\/}\\}
\vspace*{15mm}
\end{center}

\begin{abstract}
We propose a unitary matrix Chern-Simons model representing
fractional quantum Hall fluids of finite extent on the cylinder. 
A mapping between the states of the two systems is established.
Standard properties of Laughlin theory, such as
the quantization of the inverse filling fraction and of the 
quasiparticle number, are reproduced by the quantum mechanics of
the matrix model.  We also point out that this system is 
holographically described in terms of the one-dimensional
Sutherland integrable particle system.
\end{abstract}

\end{titlepage}

\section{Introduction}

Chern-Simons actions for gauge fields, since their introduction
to physics \cite{DJT}, have found numerous applications in
elementary particle and condensed matter situations.
In the last of these developments,
Chern-Simons theory on the noncommutative plane has
been proposed by Susskind as an effective
description of the fractional quantum Hall fluid \cite{Suss}.
Specifically, the ground state of this theory can be
interpreted as the Laughlin state for an infinite
number of electrons \cite{Laug}. The filling fraction corresponds
to the inverse coefficient of the Chern-Simons action.

The basic idea behind this identification is that the 
noncommutative Chern-Simons action describes a (noncommutative)
magnetic membrane, which, in turn, is equivalent to a magnetic
fluid. The connection between (commutative) membranes and fluids
is old \cite{Gol,BH} and has recently been extended to supersymmetric
fluids with spin \cite{JP}.
The new ingredient in the quantum Hall connection 
is the proposal by Susskind that an essentially noncommutative
fluid is appropriate in order to incorporate the discrete nature
of electrons. This noncommutativity exists already at the classical
level, and is distinct from the (quantum) noncommutativity
of the coordinates of particles in the lowest Landau level
\cite{GJ,DuJT} and its extension to magnetohydrodynamics 
\cite{GJPP}.

The above Chern-Simons theory can describe only an infinite
number of electrons on an infinite plane. In a previous paper
we proposed a regularized version of the noncommutative theory 
on the plane in the form of a Chern-Simons matrix model with 
boundary terms \cite{PolFQ}. This model describes a system of
finitely many electrons (a quantum Hall `droplet') and was 
shown to reproduce all the relevant physics of the
finite Laughlin states, such as boundary excitations, quantization
of the filling fraction and quantization of the charge of
quasiparticles (fractional holes). We further pointed out that
the matrix model, and thus also the quantum Hall system, is
equivalent to the Calogero model \cite{Cal}-\cite{PolLH}, 
a one-dimensional system of
particles whose connection to fractional statistics \cite{PolFS,LM},
anyons \cite{PolAN,WH,BHKV,Ouv} and the quantum Hall system 
\cite{AI,IR} has long been established.
An explicit (although non-unitary) mapping between the states of
the matrix model and Laughlin states was presented in \cite{HvR}.

It is of interest to extend the correspondence between the
noncommutative Chern-Simons matrix model and the quantum Hall
system for spaces of different topologies which have compact
dimensions. This is of both theoretical and practical significance,
since compact spaces provide a natural regularization and
have been used in alternative approaches to Laughlin states.

In this paper we shall present such a generalization for the case of
a space with cylindrical topology, in the form of a unitary matrix
model. We shall identify and analyze its classical and quantum states
and establish a mapping to Laughlin
states. Similarly to the planar case, we shall demonstrate that
this model is equivalent to a periodic version of the Calogero
model known as the Sutherland model. 
Finally we shall conclude with some directions for future research.

\section{Chern-Simons theory on the noncommutative plane
and quantum Hall states}

Before analyzing the problem on the cylinder, we will review
the basic features of Chern-Simons (CS) theory on a noncommutative
plane and a commutative time and its connection
with quantum Hall states, as proposed by Susskind \cite{Suss}.
We will also briefly review the corresponding Chern-Simons 
matrix model describing finite quantum Hall droplets \cite{PolFQ}.

\subsection{Noncommutative Chern-Simons theory from magnetic fluids}

The system to be described consists of an incompressible fluid
of $N \to \infty$ spinless
electrons on the plane in an external constant magnetic field $B$ 
(we take their charge $e=1$). 
Their coordinates are represented by two
(infinite) hermitian matrices $X_a$, $a=1,2$, that is, by two
operators on an infinite Hilbert space. The average electron density
is $\rho_0 = 1/2\pi \theta$. The action is the analog of the gauge 
action of particles in a magnetic field:
\be
S = \int dt\, \frac{B}{2} \, \Tr \left\{ \epsilon_{ab} ({\dot X}_a 
+i [A_0 , X_a ]) X_b + 2\theta A_0 \right\}
\label{CSX}
\ee
with $\Tr$ representing (matrix) trace over the Hilbert space and
$[.,.]$ representing matrix commutators. Similar actions first appeared
in matrix Chern-Simons theory as a possible approach to the 
fundamental formulation of M-theory \cite{Smol}. The specific action
above has the form of a
noncommutative CS action in the operator formulation \cite{PolCS}.
Gauge transformations are conjugations of $X_a$ by arbitrary
time-dependent unitary operators which are compact enough to leave
traces invariant. These have nontrivial topology and lead to level
quantization \cite{NP} (see \cite{Harvey} for an analysis of this class
of transformations).  In the quantum Hall context they take the meaning 
of reshuffling the labels of the electrons, a generalization of particle
permutation operators. Equivalently, the
$X_a$ can be considered as coordinates of a two-dimensional fuzzy
membrane, $2\pi\theta$ playing the role of an area quantum and gauge 
transformations realizing area preserving diffeomorphisms.

The time component of the gauge field ensures gauge invariance, 
its equation of motion imposing the Gauss law constraint
\be
-i B \, [ X_1 , X_2 ] = B \theta = \frac{1}{\nu}
\label{XX}
\ee
with $\nu = 2\pi \rho_0 / B$ being the filling fraction.
The canonical conjugate of $X_1$ is $P_2 = B X_2$, so the
operator in the left-hand side of (\ref{XX}) is  the 
generator of gauge transformations on $X_a$.
Since gauge transformations are interpreted
as reshufflings of particles, (\ref{XX}) above has the interpretation
of endowing the particles with quantum statistics of order $1/\nu$.

We will assume that $X_1 , X_2$ provide an irreducible representation
of the Gauss law (\ref{XX}), else we would be describing multiple layers
of quantum Hall fluids. This representation is essentially unique,
modulo gauge transformations, so there is a unique state in this
theory (the vacuum).
Deviations from the vacuum state can be achieved by introducing
sources in the action \cite{Suss}. A localized source at the origin 
has a density
of the form $\rho = \rho_0 - q \delta^2 (x)$ in the continuous
(commutative) case, representing a point source of particle number
$-q$, that is, a hole of charge $q$ for $q>0$. The noncommutative 
analog of such a density is
\be
[ X_1 , X_2 ] = i \theta ( 1 + q | 0 \ra \la 0 | )
\label{q}
\ee
where $|n\ra$, $n=0,1,\dots$ is an oscillator basis for the
(matrix) Hilbert space, $|0\ra$ representing a state of minimal
spread at the origin. In the membrane picture the right-hand side
of (\ref{q}) corresponds to area and implies that the area quantum
at the origin has been increased to $2\pi\theta(1+q)$, therefore
piercing a hole of area $A = 2\pi\theta q$ and creating a particle 
deficit $q = \rho_0 A$. We shall call this a quasihole state.
For $q>0$ a solution of (\ref{q}) is
\be
X_1 + i X_2 = \sqrt{2\theta} \sum_{n=1}^\infty \sqrt{n+q}
\, | n-1 \ra \la n |
\label{Xqpos}
\ee
The above assumes that $|0\ra$ is really a state at the origin,
meaning $( X_1 + i X_2 ) |0\ra = 0$. Without this residual
condition (\ref{q}) has many more solutions, since $|0\ra$ is
a gauge-dependent state that can be reshuffled around. For instance,
a class of solutions is
\be
{\tilde X}_1 + i {\tilde X}_2  = P \, \sqrt{2\theta} \sum_{n=1}^\infty 
\sqrt{n+q\vartheta(n-n_o )} \, | n-1 \ra \la n | \, P
\ee
with $\vartheta(s)$ the usual step function ($\vartheta(s) = 1$ if $s>0$,
else $\vartheta(s)=0$) and $P$ the permutator of $|0\ra$ and $| n_o \ra$
\be
P = 1 - (|0\ra - | n_o \ra)(  \la 0| - \la n_o |)
\ee
This represents an annular hole of charge $q$ at distance
$\sim \sqrt{\theta n_o}$ from the origin.

For quasiparticles ($q<0$), as long as $-q<1$ we have a similar
equation and solution. For $-q>1$, however, clearly
equations (\ref{q}) and (\ref{Xqpos}) cannot hold  
since the area quantum cannot be diminished below zero.
The correct equation is, instead,
\be
[ X_1 , X_2 ] = i \theta \left( 1 - \sum_{n=0}^{k-1} 
| n \ra \la n | - \epsilon | k \ra \la k | \right)
\label{qneg}
\ee
where $k$ and $\epsilon$ are the integer and fractional part
of the quasiparticle charge $-q$. The solution of (\ref{qneg}) is
\be
X_1 + i X_2 = \sum_{n=0}^{k-1} z_n | n \ra \la n | + 
\sqrt{2\theta} \sum_{n=k+1}^\infty \sqrt{n-k-\epsilon} 
\, | n-1 \ra \la n |
\label{Xneg}
\ee
where again we assumed that $( X_1 + i X_2 ) |k\ra =0$, so
$|k\ra$ now represents the state at the origin.
In the membrane picture, $k$ quanta of the membrane have `peeled' 
and occupy positions $z_n = x_n + i y_n$ on the plane, while the
rest of the membrane has a deficit of area at the origin equal
to $2\pi \theta \epsilon$, leading to a charge surplus $\epsilon$.
The quanta are electrons that sit on top of the continuous
charge distribution. If we want all charge density to be concentrated
at the origin, we must choose all $z_n =0$, which means that
$X_1 + i X_2$ annihilates all $|n\ra$ for $n=0,1\dots k$.

More general quasihole (particle) states, with the holes
(or fractional part of the particles) positioned at arbitrary
points on the plane can easily be constructed, but we shall not do
so here. We point out that the above particle states for integer $q$
are identical to flux solitons of noncommutative gauge theory
\cite{PolFT,GN,AGMS,HKL}.

\subsection{Finite Chern-Simons matrix model}

For a finite number of electrons $N$ we take the coordinate
$X_a$ to be finite hermitian $N \times N$ matrices.
The action (\ref{CSX}), however, and the Gauss law (\ref{XX})
are inconsistent for finite matrices, and a modified
action must be written. We take \cite{PolFQ}
\be
S = \int dt \frac{B}{2} \Tr \left\{ \epsilon_{ab} ({\dot X}_a 
+i [A_0 , X_a ]) X_b + 2\theta A_0 - \omega X_a^2 \right\} 
+ \Psi^\dagger (i {\dot \Psi} - A_0 \Psi)
\label{CSPsi}
\ee
$\Psi$ is a complex $N$-vector that 
transforms in the fundamental of the gauge group $U(N)$: 
\be
X_a \to U X_a U^{-1} ~,~~~\Psi \to U \Psi
\ee
while the term proportional to $\omega$ serves as a spatial regulator
providing a harmonic potential that keeps the electrons near the origin.

The Gauss law now reads
\be
G \equiv -iB \, [ X_1 , X_2 ] + \Psi \Psi^\dagger - B\theta =0
\label{GXX}
\ee
Taking the trace of the above equation gives
\be
\Psi^\dagger \Psi = NB \theta
\ee
The equation of motion for $\Psi$ in the $A_0 =0$ gauge is 
${\dot \Psi} =0$. So we can take it to be
\be
\Psi = \sqrt{NB} \, |v\ra
\ee
where $|v\ra$ is a constant vector of unit length. So the traceless 
part of (\ref{GXX}) reads
\be
[ X_1 , X_2 ] = i\theta \left( 1 - N |v\ra \la v| \right)
\label{XXv}
\ee
This is similar to (\ref{XX}) for the infinite plane case,
with an extra projection operator, which
is the minimal deformation of the planar result (\ref{XX})
that has a vanishing trace.  $\Psi$ clearly acts
like a boundary term, absorbing the `anomaly' of the
commutator $[X_1 , X_2 ]$, very much like the case of a boundary
field theory required to absorb the anomaly of a bulk CS field theory.

The classical states of this theory are given by the set of
matrices $A = X_1 + i X_2$ satisfying (\ref{XXv}),
and can be explicitly found \cite{PolFQ}.
The ground state is found by minimizing the potential
$V = (B\omega /2) \Tr (X_1^2 + X_2^2 ) $
while imposing the constraint (\ref{XXv}).
We obtain the solution
\be
X_1 + i X_2 = \sqrt{2\theta} \sum_{n=0}^{N-1} \sqrt{n} |n-1\ra \la n | 
~,~~~ |v\ra = | N-1 \ra
\label{APN}
\ee
This is essentially a quantum harmonic oscillator 
projected to the lowest $N$ energy eigenstates. The radius squared
matrix $R^2 = X_1^2 + X_2^2$ has a finite, equidistant spectrum. So
the above solution represents a circular quantum Hall `droplet' of radius 
~$\sqrt{2N\theta}$ and particle density $\rho_0 = N/(\pi R^2) \sim 
1/(2\pi \theta)$ as in the infinite plane case. 

Excitations of the classical ground state can be considered.
A class of such excitations are perturbations of $A=X_1 + i X_2$
generated by the infinitesimal transformation
\be
A' = A + \sum_{n=0}^{N-1} \epsilon_n (A^\dagger)^n
\label{dropex}
\ee
with $\epsilon_n$ infinitesimal complex parameters. The sum is
truncated to $N-1$ since $A^\dagger$ is an $N \times N$ matrix
and only its first $N$ powers are independent.
These map the boundary of the droplet to the new boundary 
(in polar coordinates)
\be
R' (\phi ) = \sqrt{2N\theta} + \sum_{n=-N}^N c_n e^{in\phi} 
\label{newR}
\ee
where the coefficients $c_n$ are
\be
c_n = c_{-n}^* = \frac{R^n}{2} \epsilon_{n-1}
~~(n>0),~~~~c_0 =0
\label{cc}
\ee
This is an arbitrary area-preserving deformation of the
boundary of the droplet, truncated to the lowest $N$ 
Fourier modes. The above states are, therefore, 
arbitrary area-preserving boundary excitations of the
droplet \cite{Wen,IKS,CTZ}, appropriately truncated to reflect
the finite, noncommutative nature of the system.

A second class of excitations are the analogs of quasihole and
quasiparticle states. States with a quasihole of charge $-q$ at 
the origin are of the form
\be
A = \sqrt{2\theta} \left( \sqrt{q} |N\ra \la 0| + \sum_{n=1}^{N-1} 
\sqrt{n+q} |n-1\ra \la n| \right) ~,~~~ q>0
\label{qhole}
\ee
representing a circular
droplet with a circular hole of area $2\pi\theta q$ at the origin,
that is, with a charge deficit $q$. 
Note that (\ref{qhole}) stills respects the Gauss
constraint (\ref{XXv}) (with $|v\ra = |N-1\ra$) without the
explicit introduction of any external source. 
The hole and the boundary of the droplet together cancel
the anomaly of the commutator, the outer boundary part absorbing
an amount $N+q$ and the inner (hole) boundary producing
an amount $q$. 

Fractional quasiparticle states cannot be written in this model,
reflecting the fact that such states do not belong
to the $\nu = 1/B\theta$ Laughlin state. Particle
states with an integer particle number $-q=m$ and the
extra $m$ electrons positioned outside the droplet do exist,
but we shall not write their explicit form here.

We conclude by pointing out that boundary excitations, quasiholes
and particle state can all be continuously deformed to each other,
due to the finite number of degrees of freedom of the model. 
Such transformations become
highly nonperturbative in the $N \to \infty$ limit.

\section{A model for finite number of electrons on the cylinder}

The proposed model works well for electrons on an infinite plane.
For a space representing a cylinder of radius $R$ we take one of the 
coordinates, say, $X_2$, to be periodic with period $2\pi R$. 
Clearly the above model does not take into account this periodicity
and has to be suitably modified in order to correctly 
describe the physics in the compact dimension. We shall propose
here such a model, appropriate to describing one compact dimension.

There are two routes for achieving this. The first is to write a 
matrix model on the covering space and then reduce it \cite{BFSS,Taylor}, 
leading to matrices depending on an additional continuous parameter
dual to the compact dimension, that is, a field theory. The second
is to represent the compact dimensions with unitary matrices \cite{PolUM}.
As we shall demonstrate, the two approaches turn out to be equivalent
in our case.

\subsection{The Chern-Simons unitary matrix model}

We shall begin with the second approach, which is simpler and leads
more directly to the desired model. The main point is that the
coordinate $x_2$ is not single-valued on the toroidal space and thus
is not a physical observable.
An alternative coordinate which is single-valued on the 
cylinder is the exponential $e^{i x_2 /R}$. For a noncommutative
space we define the unitary operator
\be
U = e^{i X_2 /R}
\ee
Together with the hermitian operator $X_1 \equiv X$, they parametrize
a noncommutative cylinder. The planar noncommutativity relation
for $X_1, X_2$ translates into
\be
U X U^{-1} = X + \frac{\theta}{R} 
\label{XU}
\ee
To write the Chern-Simons action on such a space we imitate again the
magnetic action of a particle with coordinates $X_i$. In the Landau
gauge $A_1 = 0$, $A_2 = -B X_1$ the lagrangian reads $-B X_1 {\dot X}_2$.
Representing ${\dot X}_2$ as $-iR U^{-1} {\dot U}$ and including
a Lagrange multiplier $A_0$ which reproduces (\ref{XU}), the full action
becomes
\be
S = \int dt\, B \, \Tr \left\{ iR \, U^{-1} ( {\dot U} 
+i [A_0 , U ]) X + \theta A_0 \right\}
\label{CSXU}
\ee
It is again expressed in terms of a covariant time derivative $D_0 U$ of
the unitary operator $U$. Note that, in the above, we have adopted
the ordering $U^{-1} {\dot U}$ for the operator representing ${\dot X}_2$. 
Had we adopted the ordering ${\dot U} U^{-1}$
we would have ended with an action involving $D_0 U \, U^{-1} X$ instead
of $U^{-1} D_0 U X$. The two actions represent identical physics, upon
redefining $X \to U X U^{-1}$ (which, upon use of the noncommutativity
relation (\ref{XU}), is simply a shift $X \to X+ (\theta /R$)).

To describe a finite system consisting of $N$ electrons we need to
take the coordinates to be finite $N \times N$ matrices. The constraint
(\ref{XU}), however, is not consistent for finite matrices, just as
in the planar case, and we need to modify the action with `boundary' 
terms that render a consistent form for the constraint.
For our purposes we take
\be
S = \int dt\, B \, \Tr \left\{ iR \, U^{-1} ( {\dot U} 
+i [A_0 , U ]) X + \theta A_0 -\frac{\omega}{2} X^2 \right\}
+ \Psi^\dagger (i {\dot \Psi} - A_0 \Psi)
\label{XUPsi}
\ee
It is similar to the planar Chern-Simons matrix model in \cite{PolFQ}.
The boundary term involves $\Psi$, a complex $N$-vector boson that 
transforms in the fundamental of the gauge group $U(N)$. 
Its role is to absorb the `anomaly' of the group
commutator $U X U^{-1} - X$, analogous to a boundary
field theory required to absorb the anomaly of a bulk CS field theory.
We also added a spatial regulator term in the form of a harmonic 
oscillator potential in the direction of the cylinder's axis.  
Since the $U$-direction is compact
we need not worry about localizing particles there, and we only want
to localize them in the infinite $X$-direction.

Before analyzing this model further, we present the alternative
derivation in terms of the covering space reduction and demonstrate
that it produces a model equivalent to the model proposed above.

\subsection{The Chern-Simons matrix field theory}

An alternative approach to deriving the desired matrix model is 
to augment the dimensionality of the hermitian matrices $X_1 , X_2$
of the planar model in \cite{PolFQ} from $N$ to $pN$ and take
$p \to \infty$. $p$ represents the copies of the cylinder on the
(planar) covering space. To ensure that the state is the same on all
copies we must impose the condition that fields in different copies
are gauge equivalent; that is, the operator which shifts copies is
a unitary (gauge) transformation. Therefore, there should exist
some unitary matrix $U$, representing shifts by one copy,
satisfying
\beqar
U X_1 U &=& X_1 ~,~~~ U X_2 U^{-1} = X_2 + 2\pi R \nonumber \\
U A_0 U^{-1} &=& A_0 ~,~~~ U \Psi = e^{i\alpha} \Psi
\eeqar
This can be explicitly realized by
parametrizing the indices $I,J$ of $(X_a)_{IJ}$, $(A_0)_{IJ}$
and $\Psi_I$ as
\be
I = i + n N ~,~~~ i = 1, \dots N ~,~~~ n = \dots -1,0,1, \dots
\ee
That is, split $X_a$ in terms of $N \times N$ blocks, with
the diagonal $(n,n)$ blocks representing the electrons on the $n$-th
copy in the covering space and the off-diagonal $(n,m)$ blocks
representing effective `interactions' of electrons between
the $n$-th and $m$-th copies on the covering space. 
Clearly the $(n,n)$ copy must be the same as the $(0,0)$ copy, only
shifted by $2\pi R n$ in the $X_2$ direction. Further, the
interactions between the $m$ and $n$ copies must only depend on
the distance between the copies $m-n$. Similarly, the Lagrange multiplier
$(A_0)_{m,n}$ must impose constraints on the $m$ and $n$ copies
that depend only on their distance $m-n$. Finally, $\Psi_n$, representing
the boundary of the state in copy $n$, must be the same for all $n$,
up to an irrelevant phase. Overall we have
\beqar
(X_1)_{m,n} &=& (X_1)_{m-n} ~,~~~
(X_2)_{m,n} = (X_2)_{m-n} + 2\pi R n \delta_{mn} \nonumber \\
(A_0)_{m,n} &=& (A_0)_{m-n} ~,~~~ \Psi_n = e^{in\alpha} \Psi
\eeqar
The unitary transformation $U$ is simply the shift $n \to n+1$.

We see that the matrices $X_a$ and $A_0$ are now parametrized in terms of
an additional integer $n$. Hermiticity of $X_a$ and $A_0$ in the original
indices $I,J$ means
\be
(X_a)_n = (X^\dagger_a)_{-n} ~,~~~ (A_0)_n = (A^\dagger_0)_{-n}
\ee
We can, therefore, define the Fourier transforms
\be
X_a (\sigma) = \sum_n (X_a)_n e^{in\sigma} ~,~~~
A_0 (\sigma) = \sum_n (A_0)_n e^{in\sigma} 
\ee
with $\sigma$ a variable with periodicity $2\pi$. $X_a (\sigma)$
and $A_0 (\sigma)$ are hermitian $\sigma$-dependent 
$N \times N$ matrices. Matrix multiplication
in the original $I,J$ indices translates into matrix
multiplication pointwise in $\sigma$, while $I$-trace translates
into $\sigma$-integration and matrix trace. It is also useful 
to define
\be
\Psi (\sigma) = \sum_n \Psi_n e^{in\sigma} = \Psi \delta(\sigma+\alpha)
\ee

We can now write the original matrix model (with a confining
harmonic potential in the $X_1$ direction)
\be
S = \int dt\, B \, \Tr \left\{ - X_1 ( {\dot X}_2
+i [A_0 , X_2 ]) + \theta A_0 -\frac{\omega}{2} X_1^2 \right\}
+ \Psi^\dagger (i {\dot \Psi} - A_0 \Psi)
\ee
in terms of the matrices $X_i (\sigma) , A_0 (\sigma)$. 
A standard calculation leads to the result
\beqar
S &=& \int dt d\sigma \, B \, \Tr \left\{ - X_1 ( {\dot X}_2 
+ i [A_0 , X_2 ] - R \partial_\sigma A_0 )  -\frac{\omega}{2} X_1^2 
+ \theta A_0 \right\} \nonumber\\
&+& \int dt \, \Psi^\dagger (i {\dot \Psi} - A_0 (\sigma=\alpha) \Psi)
\eeqar
This is nothing but $1+1$-dimensional $U(N)$ Yang-Mills theory
with a Wilson line source at $\sigma =\alpha$ and a uniform background
charge. To see this, rename $t = \sigma_0$, $R^{-1} \sigma = \sigma_1$,
$X_1 = -F/\omega$, $X_2 = A_1$. Then the above action becomes
\be
S = \int d^2 \sigma \, \frac{RB}{\omega} \, \Tr \left( F F_{01}
- \frac{1}{2} F^2 + B\omega\theta A_0 \right)
+ \int dt \, \Psi^\dagger (i {\dot \Psi} - A_0 (\sigma=\alpha) \Psi)
\ee
where we defined the field strength
\be
F_{01} = \partial_0 A_1 - \partial_1 A_0 + i [ A_0 , A_1 ]
\ee
We recognize the Yang-Mills action in the first-order formalism, 
on a circular space of radius $R^{-1}$. In addition, there is
a constant background $U(1)$ charge density $B\omega\theta$ 
and a localized
source at $\sigma_1 =\alpha$ depending on $\Psi$. The latter corresponds
to an insertion of a Wilson line in the temporal direction carrying 
the direct sum of all symmetric representations of the gauge group
$U(N)$.

To make this last point explicit, consider the temporal direction
$\sigma_0$ compact and euclidean with period $T$. This also
turns $A_0$ into $i A_0$. An appropriate
gauge transformation can render $A_0 (\sigma_0 , \alpha)$ diagonal and
independent of $\sigma_0$. The diagonal elements $(A_0)_{jj}$ (no sum
in $j$) at $\sigma_1 =\alpha$ correspond to the eigenvalues of the temporal
Wilson line element at $\sigma_1 =\alpha$, $e^{i \lambda_j}$,
specifically
\be
\left( P e^{i \int_0^T A_0 dt} \right)_{jj} = e^{i \lambda_j} 
= e^{i (A_0)_{jj} T} ~~~{\rm{(no~sum~in~}} j \rm{)}
\ee
In this gauge the components of $\Psi$ decouple and become $N$
independent bosonic harmonic oscillators with frequencies $i(A_0)_{jj}$.
Integrating out $\Psi$ produces the partition function of these $N$
oscillators which, assuming normal ordering, is
\beqar
\prod_j \frac{1}{1-e^{i (A_0)_{jj} T}} &=& \prod_j \left( 1 + e^{i\lambda_j}
+ e^{2i\lambda_j} + \cdots \right) \nonumber \\
&=& 1 + \sum_j e^{i\lambda_j} + \left( \sum_j e^{i2\lambda_j} +
\sum_{j<k} e^{i\lambda_j + i\lambda_k} \right) + \cdots
\eeqar
We recognize the terms in the last sum as the trace of the temporal
Wilson loop in the singlet, fundamental, doubly symmetric etc.
representations. The above is, then, a (bosonic) oscillator representation
of a Wilson loop element, reproducing all symmetric representations.
A discussion of arbitrary representations in the context of the planar 
CS matrix model can be found in \cite{MoP}.

It is a remarkable fact that the above theory can be reduced to a
unitary matrix model identical to the one derived in the previous section.
The details are explained in \cite{MPSYM}.
The basic point is that two-dimensional Yang-Mills theory has no
propagating modes and the only dynamical degrees of freedom are in the
nontrivial holonomy of the gauge field $A_1$ around the spatial
direction. Defining the $U(N)$ line element
\be
U[a,b] = P e^{i \int_a^{b} A_1 d\sigma_1} 
\ee
the phase space variables in the gauge $A_0 =0$ are
\be
U = U[\alpha,2\pi+\alpha] ~,~~~
X = -\frac{1}{\omega} \int_\alpha^{2\pi+\alpha} U[\alpha,\sigma] \, 
\,F U[0,\sigma]^{-1} d\sigma
\ee
In terms of these the action reduces to
\be
S = \int dt\, B \, \Tr \left\{ iR \, U^{-1} {\dot U} X 
-\frac{\omega}{2} X^2 \right\} + i \Psi^\dagger {\dot \Psi}
\ee
The Gauss law for the fields $A_1$ and $F$ involves the background
charge $\theta$ and the point source $\Psi$ at $\sigma_1 =\alpha$.
Expressed in terms of the reduced phase space variables it becomes
the constraint
\be
X - U X U^{-1} = \frac{1}{RB} \Psi \Psi^\dagger - \frac{\theta}{R}
\ee
Inserting the above constraint in the action through a new field
$A_0$ (which is now only a function of time) we recover the unitary
Chern-Simons matrix model (\ref{XUPsi}). Thus, both routes to 
compactifying $X_2$ lead to the proposed unitary matrix model.

\section{Classical states of the unitary matrix model}
\subsection{General solution}

To get a feeling of the physics of the above model we shall
analyze its classical structure and states.
We can again impose the $A_0$ equation of motion as a Gauss
constraint and then put $A_0 =0$. In our case it reads
\be
G \equiv RB \, ( U X U^{-1} - X) + \Psi \Psi^\dagger - B\theta =0
\label{G}
\ee
The trace of the above equation gives as in the planar case
\be
\Psi^\dagger \Psi = NB \theta
\ee
In the $A_0 =0$ gauge $\Psi$ is constant, take it $\sqrt{NB} \, |v\ra$
with $|v\ra$ a constant unit vector. The traceless 
part of (\ref{G}) reads
\be
U X U^{-1} - X = \frac{\theta}{R} \left( 1 - N |v\ra \la v| \right)
\label{XUv}
\ee
This is similar to (\ref{XU}) for the infinite cylinder,
with an extra projection operator. Again, using the residual 
time-independent $U(N)$ gauge freedom to rotate
$|v\ra$ to the form $|v\ra = (0,\dots 0,1)$ we obtain the form 
$(\theta/R) \,{diag}\, (1,\dots, 1, 1-N)$ for the above constraint,
which is the minimal deformation of the result (\ref{XU})
that has a vanishing trace. 

The equations of motion for $X$ and $U$ read
\be
{\dot X} + [ U^{-1} {\dot U} , X ] = 0 ~,~~~
iR U^{-1} {\dot U} - \omega X = 0
\ee
or
\be
{\dot X} = i\frac{R}{\omega} \frac{d}{dt} ( U^{-1} {\dot U} ) = 0
\label{eom}\ee
This represents free motion on the manifold $U(N)$, as represented
by the unitary matrix $U(t)$, with $X$ playing the role of matrix 
momentum. It is solved by
\be
U(t) = U_0 e^{-i\omega X_0 t /R} ~,~~~ X(t) = X_0
\ee
where the constant matrices $X_0 , U_0$ satisfy the constraint 
(\ref{XUv}). We can find all the classical states of this model
by diagonalizing $U_0 = diag\, \{e^{i \phi_n}\}$. By examining the
diagonal and off-diagonal elements of (\ref{XUv}) we see that
the components of $|v\ra$ have length $|v_n |^2 = 1/N$. Choosing
their phases as $v_n = exp(-i\phi_n /2) /\sqrt{N}$, as we can do
using the residual $U(1)^N$ gauge invariance, we obtain
\be
(U_0 )_{mn} = e^{i\phi_n} \delta_{mn} ~,~~~
(X_0 )_{mn} = x_n \delta_{mn} + \frac{i\theta}
{2R \sin (\frac{\phi_m - \phi_n}{2})} (1-\delta_{mn} )
\label{XUsol}
\ee
The solution is parametrized by the $N$ compact eigenvalues of
$U_0$,  $\phi_n$, and the $N$ diagonal elements of $X$, $x_n$,
that is, by $N$ coordinates on the cylinder.

\subsection{Classical ground state}

The lowest energy state, that is, the state most closely packed
around $X_1 =0$, is found by minimizing the potential $(B\omega/2)
X^2$ while respecting the constraint (\ref{XUv}).
Implementing it with a matrix Lagrange multiplier $\Lambda$
we obtain
\be
B\omega X + U^{-1} \Lambda U - \Lambda = 0 ~,~~~
[ X , U^{-1} \Lambda U ] = 0
\label{XUL}
\ee
The above is solved by
\beqar
U_{mn} &=& e^{i\phi_0} \delta_{m,n+1} ~,~~~
X_{mn} = \frac{\theta}{R} \left( \frac{N+1}{2} - n \right) \delta_{mn} \\
\Lambda_{mn} &=& \frac{\theta B\omega}{2R} 
\left( \frac{N+1}{2} - n \right) \left( \frac{N+1}{2} - n +1\right) \delta_{mn}
~,~~~ v_n = \delta_{n,1}
\label{grst}
\eeqar

The eigenvalues of $U$ are $exp(i\phi_0 + i2\pi n/N)$ and are
evenly distributed on the compact dimension of the cylinder.
Similarly, the eigenvalues of $X$ are evenly distributed along
the axis of the cylinder and span a length $\sim N\theta /R$.
Therefore, the above solution represents a tubular quantum Hall
droplet around the cylinder with an area $A = 2\pi R \cdot
N \theta /R$ and an average density $\rho_0 = N/A = 1/(2\pi \theta)$.

We point out that the average distance between successive electrons
in the $x_1$ direction is $\theta/R$. In the planar case, particles
were evenly distributed on the plane with a density $1/2\pi\theta$
and thus an average distance of order $\sqrt\theta$. This is a
signal that, on the cylinder, quantum Hall states do not have
a constant density. In the extreme case $\theta \gg R^2$ the
electrons will behave more like one-dimensional particles with
well-defined positions along the length of the cylinder.
We also point out the existence of $\phi_0$
in the solution for $U$, which does not affect the energy.

\subsection{Equivalence to the Sutherland model}

Just as in the planar case, the matrix model above is equivalent
to a one-dimensional particle system, the so-called Sutherland model
\cite{Sut}.
This is an integrable system of $N$ nonrelativistic particles on the 
circle with coordinates and momenta $(\phi_n , p_n )$ and hamiltonian
\be
H = \sum_{n=1}^N \frac{\omega}{2B} p_n^2 
+ \sum_{n\neq m} \frac{\nu^{-2}}{4\sin^2 \frac{\phi_n - \phi_m}{2} }
\label{HSut}
\ee 
It can be thought of as the Calogero model of particles on the
line with inverse-square mutual potentials, rendered periodic in
space with period $2\pi R$.
In terms of the parameters of the model, the mass of the particles
is $B/\omega$ and the coupling constant of the two-body inverse-square
potential is $\nu^{-2}$. We refer the reader to \cite{OP,PolLH,PolMM}
for details of the derivation of the connection between the matrix
model and the Sutherland model and to \cite{MPSYM,GoN} for the connection
between 2-dimensional Yang-Mills and the Sutherland model. 
Here we simply state the relevant
results and give their connection to quantum Hall quantities.

The positions of the Sutherland particles on the circle $\phi_n$ are
the eigenvalues of $U$, while the momenta $p_n$ are the diagonal
elements of $X_1$, specifically $p_n = B x_n$. The motion of the
$\phi_n$ generated by the hamiltonian (\ref{HSut}) is compatible with
the evolution of the eigenvalues of $U$ as it evolves in time
according to (\ref{eom}). So the Sutherland model gives a 
holographic description of the quantum Hall state by monitoring
the effective electron coordinates along $X_2$, that is, the eigenvalues
of $U$.

The hamiltonians of the Sutherland and matrix model are equal
and energy states map.
The ground state is obtained by putting the particles at their
static equilibrium positions. Because of their repulsion and the
symmetry of the problem, they will form a uniform lattice of points
on the circle, as in (\ref{grst}). $\phi_0$ is the coordinate
of the center of mass of the particles. 
Sound waves on this lattice correspond to small perturbations of
the quantum Hall state. As the amplitude grows large,
they become nonlinear nondispersive waves corresponding to holes
forming in the quantum Hall particle distribution along the 
$X$-direction. In the limit they become solitons, representing
isolated particles off the quantum Hall ground state \cite{PolWS}. 
Further connections at the quantum level will be described in
subsequent sections.

\section{Quantization of the cylindrical matrix Chern-Simons model}
\subsection{Gauss law and quantization of the filling fraction}

We now proceed with the quantization of the above model. We use
double brackets for quantum commutators and double kets for quantum
states to distinguish them from matrix commutators and $N$-vectors.

The 
canonical structure of (\ref{XUPsi}) implies the Poisson brackets
\be
\{ X_{ij} , U_{kl} \} = \frac{i}{iBR} \delta_{il} U_{kj}
\ee
This means that $-BRX$ is the generator of right-rotations of
the matrix $U$. Indeed, for any  hermitian matrix $\epsilon$
\be
\{ -BR\, \tr(\epsilon X) , U \} = i U \epsilon
\ee
Quantum mechanically $X$ should also generate right-rotations
of $U$. In the $U$-representation, then, it acquires the form
\be
X_{ij} = -\frac{1}{BR} \, U_{kj} \frac{\partial}{\partial U_{ki}}
\ee
As a result, ${\cal R} \equiv -BRX$ satisfies the $U(N)$ algebra:
\be
[[ {\cal R}_{ij} , {\cal R}_{kl} ]] = \delta_{il} {\cal R}_{kj} 
- \delta_{kj} {\cal R}_{il}
\label{UN}
\ee
with the obvious hermiticity condition $X_{ij}^\dagger = X_{ji}$.
Similarly, the classical matrix ${\cal L} \equiv BR \, U X U^{-1}$ 
generates left-rotations of $U$. It should therefore be ordered as
\be
{\cal L}_{ij} = -U_{ik} \frac{\partial}{\partial U_{jk}}
\ee
and also satisfies the $U(N)$ algebra (\ref{UN}). 
The sum of these operators $G_U \equiv {\cal L} + {\cal R}$ 
satisfies the $SU(N)$ algebra ($\cal L$ and $\cal R$ have equal
and opposite $U(1)$ parts) and generates unitary conjugations
of the matrix $U$.

The components of $\Psi$ are harmonic oscillators, satisfying
\be
[[ \Psi_i , \Psi_j^\dagger ]] = \delta_{ij}
\ee
The matrix $G_\Psi \equiv \Psi \Psi^\dagger$ generates rotations 
of the vector $\Psi$, and must also satisfy the $U(N)$ algebra 
(\ref{UN}).  It should therefore be ordered as
\be
(G_\Psi )_{ij} = \Psi_j^\dagger \Psi_i
\ee

The Gauss law constraint then acquires the form
\be
( {\cal L} + {\cal R} + G_\Psi - \theta B ) |phys \ra \ra = 0
\ee
The situation is similar to the planar case. $G_U$ contains only
symmetric products of the adjoint, with a number of boxes in their
Young tableau ($Z_N$ charge) a multiple of $N$, $kN$ ($k$ integer).
$G_\Psi$ contains only totally symmetric representations with $U(1)$
charge equal to the number of boxes in the Young tableau. 
For the total representation to be in the singlet, as required by 
the Gauss law, $G_U$ and $G_\Psi$ must be in conjugate representations
and thus $G_\Psi$ must also have a number of boxes $kN$. Moreover,
the trace ($U(1)$ charge) of $G_\Psi$ must cancel $N\theta B$. So
we obtain the quantization condition
\be
B \theta = k ~,~~~ k={\rm integer}
\label{quc}
\ee
Again, this is related to the level quantization of the noncommutative
Chern-Simons action \cite{NP,BLP} and can also be attributed to a
global gauge anomaly of the model \cite{PolMM}.
The above condition will lead to the quantization of the inverse filling
fraction, as in Laughlin theory. We anticipate the actual result as
$\nu = 1/(k+1) \equiv 1/n$; the shift from $k$ to $k+1$ is a quantum
correction.

\subsection{Quantum states}

The hamiltonian is
\be
H = \frac{B\omega}{2} \, \tr X^2 = \frac{\omega}{2BR^2} \, 
\tr {\cal L}^2 = \frac{\omega}{2BR^2} \, \tr {\cal R}^2
\ee
It is the Laplacian on the group manifold $U(N)$
(remember that $X$ is essentially the momentum of $U$), and is
proportional to the common quadratic Casimir of $\cal L$ or $\cal R$.

Since the space $U(N)$ is curved, we could add to the hamiltonian
a term proportional to the curvature. Such terms can always arise
through quantum ordering effects; a particular value, equal to
$1/8$ times the curvature, is singled out from conformal invariance.
In our case the curvature is constant and, as we shall see, the
addition of a constant term as above will make the spectrum
especially simple and suggestive.

Quantum states can be represented in terms of wavefunctions of $U$.
A particular set of such wavefunctions diagonalizes $H$. Specifically,
consider the matrix elements $R_{\alpha \beta} (U)$ of the matrix $U$ 
in some irreducible representation of $SU(N)$ $R$ of dimension $d_R$.
By Schur's lemma, any arbitrary function of $U$ can be expanded in
terms of the above functions. 

Each matrix element $R_{\alpha \beta}$ above is, in fact,
an eigenstate of $H$. To see this, note that under arbitrary left-
and right-rotations of $U$ the states transform as
\be
R_{\alpha \beta} (V^{-1} U W ) = R_{\alpha \gamma}^{-1} (V)
R_{\gamma \delta} (U) R_{\delta \beta} (W)
\ee
So the multiplet $R_{\alpha \beta} (U)$, $\alpha,\beta = 1, \dots d_R$
 transforms in the $R$
representation under right-rotations of $U$ and in the conjugate 
$\bar R$ representations under left-rotations of $U$. Since $H$
is the quadratic Casimir of $\cal R$ or $\cal L$ we obtain
\be
H R_{\alpha \beta} (U) = 
\frac{\omega}{2BR^2} \, C_{2,R} \, R_{\alpha \beta} (U)
= E_R \, R_{\alpha \beta} (U) 
\ee
So the spectrum of $H$ consists of all quadratic Casimirs $C_{2,R}$,
each with a degeneracy $d_R^2$ corresponding to the different matrix
elements of $R_{\alpha \beta} (U)$.

We still need to impose the Gauss law constraint. According to the
discussion of the previous section, it stipulates that the states for $U$
transform in a totally symmetric representation $S_k$, with $kN$ boxes,
under conjugations of $U$. This means that we must pick the corresponding
representation for $G_U = {\cal L} + {\cal R}$. Clearly the $d_R^2$
states $R_{\alpha \beta} (U)$ transform in the $R \times {\bar R}$
representation under $G_U$. We must, therefore, decompose 
$R \times {\bar R}$ into
irreducible components and pick the symmetric representation $S_k$
among the components. Each $S_k$ corresponds to a unique physical
state (the components within $S_k$ are contracted in a unique way
with the components of the $\Psi$-representation $S_k$ to give a
singlet). This fixes the degeneracy of the eigenvalue $E_R$ to the
number of times $D_{R,k}$ that $S_k$ appears in $R \times {\bar R}$.

The above takes care of the $SU(N)$ part of the wavefunction. We can
always assign an arbitrary $U(1)$ part by multiplying the wavefunction
with $(\det U)^q$. If we want single-valuedness under the transformation
$U \to exp(i 2\pi) \, U$, which corresponds to rotations around the
cylinder, we must have $q = p/N$ with $p$ an integer, corresponding to
integer $U(1)$-charge of the state under $\cal L$ or $\cal R$. Since
the $U(1)$ charge of $R$ is the number of boxes, we can simply take
the $N$ rows of its Young tableau to have either positive or negative
length. 

We have reduced the physical spectrum of the model to pure group
theory. In the present case, however, we can do better than that.
Using standard Young tableau multiplication rules, it can be verified
that in order to obtain  $S_k$ in the product 
$R \times {\bar R}$, the representation $R$ must have lengths of 
Young tableau rows $\ell_j$ that satisfy
\be
\ell_j - \ell_{j+1} \geq k ~,~~~ j=1, \dots N
\label{ell}
\ee
For each such $R$, $S_k$ is contained exactly once in $R \times {\bar R}$.

The next interesting fact is that the quadratic Casimir of $U(N)$ can
be expressed in terms of the spectrum of free fermions on the circle
\cite{Nambu}.  Specifically, define the fermion `momenta'
\be
p_j = \ell_j + \frac{N+1}{2} -j ~,~~~ j=1, \dots N
\label{pj}
\ee
which satisfy $p_j > p_{j+1}$. Then the expression for $C_{2,R}$ is
\be
C_{2,R} = \sum_{j=1}^N \left( p_j^2 - p_{j,0}^2 \right)
\ee
where $p_{j,0}$ are the `ground state' momenta, corresponding to
the singlet representation $\ell_j =0$:
\be
p_{j,0} = \frac{N+1}{2} -j 
\ee

The final observation is that if we add to the hamiltonian a curvature
term with coefficient $1/8$, as mentioned in the beginning of the section,
the `ground state' term in $C_{2,R}$ exactly cancels. We are left,
then, with the spectrum of free identical particles on the circle, but with
an enhanced exclusion principle. Specifically
\be
E_R = \sum_j E_j = \frac{\omega}{2BR^2} \sum_j p_j^2
\ee
where $E_j$ are effective single-particle pseudo-energies, and the
$p_j$ are single-particle pseudomomenta.  Because of (\ref{pj}) 
and (\ref{ell}), the $p_j$ obey
\be
p_j - p_{j+1} \geq k+1 = n
\ee
We obtain the spectrum of free nonrelativistic particles of mass
$\omega/B$ on a circle of radius $R$, obeying exclusion statistics
of order $n=k+1$. This is the spectrum of Sutherland particles,
which we have recovered entirely in the matrix model context.

It is reasonable to interpret $E_j$ as the quantum analogs of the
eigenvalues of the potential $(B\omega/2) X^2$. This means that the
positions of the electrons along the $X$-direction are 
\be
x_j = \frac{1}{BR} \, p_j
\ee
The ground state quasimomenta 
\be
p_{j,gs} = n \frac{N+1-2j}{2}
\ee
form a `Fermi sea' with distance $n$ between successive momenta. 
The $x_j$, then, correspond to evenly spaced electrons
with a distance $d = n/BR$ between them, and therefore an average
density $\rho = 1/(2\pi R \cdot d)$. The filling fraction then is
\be
\nu = \frac{2\pi \rho}{B} = \frac{1}{n}
\ee
justifying the interpretation of $n=k+1$ as the quantized inverse
filling fraction. $k=0$ (the singlet sector) corresponds to free
fermions, reproducing the fully filled $n=1$ Landau level.

Quasiparticle and quasihole states are identified in a way 
completely analogous to the planar case.
A quasiparticle state is obtained by peeling a `particle' from
the surface of the sea (quasimomentum $p_{1,gs}$) and putting it 
to a higher value $p_1 > n(N-1)/2$. This corresponds to an electron
at position $x \sim p_1 /B$ along $X$
in a state covariant under rotations of the cylinder. 

Quasiholes correspond to the minimal excitations of the ground
state inside the quantum Hall tubular distribution. 
This is achieved by leaving all quasimomenta $p_j$
for $j \geq r$ unchanged, for some integer $r$, and increasing all 
$p_j$, $j< r$ by one unit:
\beqar
p_j &=& n\frac{N+1-2j} ~,~~~~~~~ j \geq r \\
    &=& n\frac{N+1-2j}{2} +1 ~~~  j < r 
\eeqar
This increases the gap between $p_r$ and $p_{r+1}$ to $n+1$
and creates a minimal `hole' at position $x \sim p_{r,gs} /BR$.
As in the planar case, removal of one particle corresponds to
the creation of $n$ holes, and therefore the particle number 
of the hole is $-q = -1/n = -\nu$.  We again recover
the quantization of the quasihole charge in fundamental units of
\be
q_h = \nu = \frac{1}{n}
\label{holeq}
\ee
in accordance with Laughlin theory.

Finally, we have center-of-mass ($U(1)$) excitations, achieved
by shifting all $p_j$ by the same amount. This corresponds to
translations of the electron state along the axis of the cylinder.

\section{Correspondence to Laughlin states}

The discussion in the last section demonstrates that there is
a qualitative mapping between Laughlin states and matrix states.
It is desirable to establish a more precise and explicit mapping
between the two systems. 

The wavefunction of the electrons in the lowest Landau level is a
function of two variables. Alternatively, we can use the reduced
phase space representation, in which spatial coordinates do not
commute and span a quantum phase space. The mapping between the
two representation in the plane is through coherent states.
Laughlin states can then be considered as particular restricted
many-body Landau wavefunctions, or corresponding states in the
reduced phase space.

In order to establish the mapping between matrix and Laughlin
states we will first define coherent states on a cylindrical
phase space, then establish the correspondence with Laughlin
states on the cylinder and finally map to matrix model states.

\subsection{Coherent states on a cylindrical phase space}

On a planar quantum phase space with coordinates $X,Y$ satisfying
$[ X, Y ] =i/B$ we can define creation and annihilation operators
$a,a^\dagger$ as $a = (X + i Y)\sqrt {B/2}$. Coherent states, then,
are defined as eigenstates of the annihilation operator $a$ and
represent states of minimum uncertainty.

On a cylindrical phase space with one compact coordinate, say,
$Y \equiv Y + 2\pi R$, the operator $Y$ is multivalued and thus
unphysical. The above creation-annihilation operators are therefore
unphysical and we cannot use them to define coherent states.
The operator $exp[(X+iY)/R ]$, however, is single-valued and physical.
We will then define coherent states $|z\ra$ by the relation
\be
e^{(X+iY)/R} |z\ra = e^{z/R} |z\ra
\ee

We can easily find the expression for the wavefunction of $|z\ra$.
In the $Y$ representation, wavefunctions are periodic functions
of $Y$ and can be expanded in terms of the usual momentum states $|n\ra$:
\be
\la Y | n \ra = e^{inY/R} ~,~~~  n= \dots ,-1,0,1,\dots
\ee
In this representation $X$ acts as $(i/B) \partial / \partial Y$.
A straightforward calculation shows that $|z\ra$ is of the form
\be
|z\ra = N \sum_n e^{-\frac{nz}{R} - \frac{n^2}{2BR^2}} \, |n\ra
\ee
where $N$ is a normalization factor.
Writing $z=x+iy$, we see that $|z\ra$ is a state with $X$
centered around $x$ (although $X$ has discrete eigenvalues)
and $Y$ centered around $y$ (modulo $2\pi R$).

It is convenient to choose the normalization $N$ as 
\be
N = (2 \pi^{3/2}B^{1/2} R)^{-1/2} e^{-B x^2 /2}
= N_o e^{-B x^2 /2}
\ee
Then it is straightforward to verify the completeness relation
\be
\int dz d{\bar z} \, |z \ra \la z | = 1
\ee
Finally, the coherent state wavefunction of an $n$-state is
\be
\la z |n\ra = N_o \, e^{i\frac{ny}{R} - \frac{(n+BRx)^2}{2BR^2}}
\ee
or, defining $w = exp(-x+iy)$ as the corresponding analytic 
coordinate on the cylinder,
\be
\la w |n\ra = N_o \, w^n e^{-\frac{B}{2} x^2 -\frac{n^2}{2BR^2}}
\ee

\subsection{Laughlin states on the cylinder}

We start by presenting the single-particle wavefunctions for the
lowest Landau level on the cylinder. These are well-known, and
can easily be found in the Landau gauge $A_x =0$, $A_y = -Bx$.
In a basis diagonalizing the momentum in the $Y$-direction they
become
\be
\la x,y |n\ra_L \equiv \psi_n (z=x+iy) = 
e^{i\frac{ny}{R} - \frac{B}{2} \left( x + \frac{n}{BR} \right)^2 } =
w^n e^{-\frac{B}{2} x^2 -\frac{n^2}{2BR^2}}
\label{LL}
\ee
They are `stripe' states, exponential in the $y$-direction
and gaussian in the $x$-direction with a center shifted by
$-n/BR$. 

We immediately see the similarity with the coherent states in the
reduced phase space. We conclude that Landau wavefunctions in terms
of coordinates on the cylinder equal the corresponding coherent
states on the reduced phase space.

$N$-body Laughlin states on the cylinder can be defined in a way 
analogous to the plane. There, the electrons were restricted to states
containing the `ground state' factor
\be
\psi_{gs} = \prod_{j<k} ( z_j - z_k )^n e^{-\frac{B}{2} \sum_j
| z_j |^2}
\label{VP}
\ee
For $n=1$ this is the Slater determinant of the lowest $N$ angular momentum
eigenstates, which reduces to the Vandermonde determinant for the variables
$z_i$ times the (non-analytic) N-body oscillator ground state. For higher $n$
the analytic (Vandermonde) part is raised to the $n$-th power, while the
ground state part remains the same.

A similar construction can
be repeated on the cylinder, but with the momentum around the
cylinder replacing the angular momentum. The corresponding
Laughlin wavefunction would be
\be
\psi_{gs} = \prod_{j<k} ( w_j - w_k )^n e^{-\frac{B}{2} \sum_j
x_j^2}
\ee
The relevant single-particle states appearing in (\ref{CP}) are, now,
(\ref{LL}). The lowest $y$-momentum appearing in (\ref{CP}) is 0,
while the highes is $n(N-1)$, corresponding to positions in the 
$x$-direction from 0 to $-n(N-1)/BR$. So this correspond to a tubular
Hall state with  area $A \sim 2\pi n (N-1) /B$, corresponding to a filling 
fraction $\nu = 1/n$. 

To minimize the potential in $x$, the state should be centered around
$x=0$ and thus the above state
should be shifted to the right by $\Delta x = n(N-1)/2BR$. This 
amounts to multiplying the wavefunction by the shift factor
$\prod_j w_j^{-n(N-1)/2}$.
The resulting, properly centered Laughlin wavefunction is
\be
\psi_{gs} = \prod_{j<k} ( u_{jk} )^n e^{-\frac{B}{2} \sum_j
x_j^2} ~,~~~ u_{jk} = \left( \frac{w_j}{w_k}\right)^\frac{1}{2}
- \left( \frac{w_k}{w_j}\right)^\frac{1}{2}
\label{CP}
\ee
For small distances the above wavefunction has the same behavior
as the planar Laughlin state. It is interesting that its probability density 
also admits a plasma interpretation, as in the planar case. Indeed, the
Coulomb potential on a torus takes the form
\be
V(x,y) = \half \ln \left(\sinh \frac{z}{2R}  \sinh \frac{\bar z}{2R} \right)
= \half \ln \left( w^\half - w^{-\half} \right) +\half \ln \left( {\bar w}^\half 
- {\bar w}^{-\half} \right)
\ee
Therefore, the Vandermonde part of $|\psi_{gs} |^2$ in (\ref{CP}) will 
reproduce the exponential of the mutual Coulomb potential of particles
on the torus, while the gaussian part represents the potential of a
`neutralizing' constant background charge distribution, in exact analogy
to the planar case.

The above ground state wavefunction can just as well be interpreted 
as the coherent wavefunction on the reduced phase space.
Excited states above the ground state can be written by multiplying
this wavefunction by any totally symmetric wavefunction of the $w_i$.
Such wavefunctions can be uniquely expressed in terms of the Schur
basis
\be
S_{c_1 , c_2 , \dots} = \left( \sum_j w_j^{c_i} \right)
\left( \sum_k w_k^{c_2} \right) \cdots
\ee
and the general state will be a product of $S_{c_1 , c_2 \dots}$
and $\psi_{gs}$:
\be
\psi_{c_1 , c_2 \dots } = \left( \sum_j w_j^{c_i} \right)
\left( \sum_k w_k^{c_2} \right) \cdots
\prod_{j<k} ( w_j - w_k )^n e^{-\frac{B}{2} \sum_j x_j^2} 
\label{Lau}
\ee
where we omitted the shift factor since, as all symmetric functions,
it can be reproduced in terms of Schur functions.

\subsection{Mapping to matrix states}

The states of the matrix model can be explicitly written in a
way analogous to the one for the plane \cite{HvR}. We will work
in the $U$ representation for the matrices and the oscillator
representation for the $\Psi$. Define a ground state wavefunction
$|0\ra$ which is the Fock vacuum of the oscillators $\Psi$ and
the singlet (constant) in $U$; that is, 
\be
\Psi_j |0\ra = X_{jk} |0\ra = 0
\ee
Excited states can be obtained by applying $\Psi^\dagger$'s and
$U$'s on $|0\ra$. For the resulting state to be gauge invariant,
all indices of $U_{jk}$ and $\Psi_j^\dagger$ must be contracted,
either with each other or with the $SU(N)$ antisymmetric tensor
$\epsilon^{j_1 j_2 \dots j_N}$.

The $U(1)$ gauge constraint (\ref{quc}), on the other hand,
stipulates that each physical state should have exactly $kN$
operators $\Psi^\dagger$. The minimal way that we can contract the
indices of the above $\Psi^\dagger$'s is
\be
|gs\ra = \left( \epsilon^{j_1 \dots j_N} \Psi_{j_1}^\dagger 
(\Psi^\dagger U)_{j_2} \cdots (\Psi^\dagger U^{N-1} )_{j_N} 
\right)^k |0\ra
\ee
This can be considered the ground state of the system. The powers
of $U$ that appear must clearly be all different, and we chose
them to span the values $0,1,\dots N-1$. We could have, instead,
chosen the values $-\frac{N-1}{2} , \dots \frac{N-1}{2}$, which would
have given the $U$ part of the state a vanishing $U(1)$ charge.
This can always be done a posteriori by multiplying with
$\det U^{-(N-1)/2}$, and we shall stay with $|gs\ra$ above for
simplicity.

Other states can be obtained by multiplying with gauge invariant
combinations of $U$'s (no more $\Psi^\dagger$ are allowed). These
are spanned by the Schur functions
\be
S_{c_1 , c_2 \dots} = \tr U^{c_1} \tr U^{c_2} \cdots
\ee
and a complete basis for the matrix states is
\be
| c_1 , c_2 \dots \ra = \tr U^{c_1} \tr U^{c_2} \cdots
\left( \epsilon^{j_1 \dots j_N} \Psi_{j_1}^\dagger 
(\Psi^\dagger U)_{j_2} \cdots (\Psi^\dagger U^{N-1} )_{j_N} 
\right)^k |0\ra
\label{ccc}
\ee

The next step is to parametrize $U$ in terms of diagonal and
angular variables:
\be
U = V^{-1} \Lambda V
\ee
with $V$ an $SU(N)$ matrix and $\Lambda = diag \{ e^{i \phi_j } \}
\equiv diag \{ W_j \}$
the eigenvalues of $U$. Each $\epsilon$-factor in the states (\ref{ccc})
becomes
\beqar
&& \epsilon^{j_1 \dots j_N} \Psi_{j_1}^\dagger 
(\Psi^\dagger U)_{j_2} \cdots (\Psi^\dagger U^{N-1} )_{j_N} 
\nonumber \\
&=& \epsilon^{j_1 \dots j_N} (\Psi^\dagger V^{-1})_{k_1} V_{k_1 j_1} 
(\Psi^\dagger V^{-1} )_{k_2} W_{k_2} V_{k_2 j_2} \cdots 
(\Psi^\dagger V^{-1} )_{k_N} W_{k_N}^{N-1} V_{k_N j_N}
\nonumber \\
&=& \left\{ \epsilon^{j_1 \dots j_N} V_{k_1 j_1}  \cdots V_{k_N j_N} 
\right\} 
\left\{ W_{k_1}^0  \cdots W_{k_N}^{N-1} \right\}
\left\{
(\Psi^\dagger V^{-1})_{k_1}  \cdots (\Psi^\dagger V^{-1} )_{k_N}
\right\} 
\label{eVW}
\eeqar
Since the $\epsilon$ tensor antisymmetrizes the indices $j_n$,
the indices $k_n$ appearing in $V_{k_n j_n}$ in the first bracket
are also antisymmetrized and we obtain
\be
\epsilon^{j_1 \dots j_N} V_{k_1 j_1}  V_{k_2 j_2} \cdots 
V_{k_N j_N} = \epsilon^{k_1 \dots k_N} \det V
= \epsilon^{k_1 \dots k_N} 
\ee
($\det V =1$ since $V$ is special unitary). Defining
\be
\chi = V \Psi 
\ee
the expression (\ref{eVW}) becomes
\be
\epsilon^{k_1 \dots k_N} \, W_{k_1}^0 \cdots W_{k_N}^{N-1} \,
\chi^\dagger_{k_1} \cdots \chi^\dagger_{k_N} 
\ee
Since all $k_n$ are distinct, the product of the $\chi_{k_n}^\dagger$ 
is simply $\chi_1^\dagger \chi_2^\dagger \cdots \chi_N^\dagger$. 
The remaining $W$'s with the
$\epsilon$ symbol reproduce the Vandermonde determinant. So the
expression above becomes
\be
\prod_{j<m} ( W_j - W_m) \prod_j \chi^\dagger_j
\ee
The Schur functions, on the other hand, become
\be
\tr U^c = \sum_j W_j^c
\ee
Overall, the states of the matrix model (\ref{ccc}) take the form
\be 
| c_1 , c_2 \dots \ra = 
\left( \sum_j W_j^{c_1} \right) \left( \sum_j W_j^{c_2} \right)
\cdots \prod_{j<m} ( W_j - W_m)^k \left( \prod_j \chi^\dagger_j
\right)^k |0\ra
\label{ccWx}
\ee

The operators $\chi_j$ defined above are also harmonic oscillators
and they satisfy $[[ \chi_j , \chi_k^\dagger ]] = \delta_{jk}$. So the 
oscillator state
\be
|\Omega\ra = \left( \prod_j \chi^\dagger_j \right)^k |0\ra
\ee
appearing above has a norm $\la \Omega | \Omega \ra$ independent of
the matrix $V$ which enters in the definition the $\chi_j$. 
Since no other oscillator
state can ever appear, the above state, used for calculations of
matrix elements, is effectively independent of $V$ and $W_j$.
The matrix $V$ has, therefore, completely disappeared from the
picture.

When calculating matrix elements between states, we must integrate 
with the Haar measure $[dU]$ over the matrix $U$. This is
\be
[dU] = [dV] \, \prod_{j<k} | W_j - W_k |^2 \, \prod_j d\phi_j
\ee
The integration $[dV]$ over $V$ produces a constant, since
nothing depends on $V$. We are left with an integration over the
eigenvalues $\phi_j$, with an additional Vandermonde term coming
from the measure. It is convenient to incorporate this measure
into the definition of the states, so that matrix elements can be
calculated with a flat measure over the $\phi_j$. This introduces
an additional power of the Vandermonde determinant in (\ref{ccWx}),
shifting $k$ to $k+1 =n$. This is the origin of the renormalization
of the filling fraction that we mentioned before. The final matrix
states are, therefore,
\be
| c_1 , c_2 \dots \ra = 
\left( \sum_j W_j^{c_1} \right) \left( \sum_j W_j^{c_2} \right)
\cdots \prod_{j<m} ( W_j - W_m)^n |\Omega\ra
\label{Ust}
\ee
These are identical in form to the corresponding Laughlin states
(\ref{Lau}) upon mapping $W_j = e^{i\phi_j}$ to $w_j$ and identifying
the ground state $|0\ra$ with the bosonic gaussian factor in (\ref{Lau}).

The above provides a formal mapping between the states of the
unitary Chern-Simons matrix model and Laughlin states on the
cylinder, much like the one for the plane \cite{HvR}. 
It should be stressed, however, that the above mapping
is {\it not} unitary. Indeed, the wavefunctions (\ref{Ust}) above
are integrated with a flat metric in $\phi_j$, while the Laughlin
states (\ref{Lau}) are integrated with the planar measure $dx dy$.
Therefore, the norms and scalar products of the two sets of states
are, in general, different. This mapping is, therefore, at best a 
qualitative one. The exact, unitary mapping between matrix 
model states and Laughlin states, planar or cylindrical, 
is still lacking. Some relevant results in the planar case are derived
in \cite{KS}.

Note, further, that the above states (\ref{Ust}) are not eigenstates
of the eigenstates of the matrix model hamiltonian $\tr X^2$. 
Eigenstates can always be constructed by forming appropriate
linear combinations of states (\ref{Ust}) with the same degree of
homogeneity, it essentially amounts to constructing the characters
of the appropriate allowed representations of $U(N)$ (see the
discussion in section 5.2).  

\section{Outlook}

We have extended a previous proposal and presented a
unitary matrix Chern-Simons model describing fractional quantum 
Hall states of $N$ electrons on the cylinder. The correspondence
of the two systems was established as a map between states 
in each Hilbert space. As in the planar case,
the quantization of the inverse filling fraction
and of the quasihole charge are straightforward consequences of
the quantum mechanics of this model. We also stressed that
the classical value of the inverse filling fraction is shifted
quantum mechanically by one unit. This can be equivalently viewed
as a renormalization of the Chern-Simons coefficient, as a
group-theoretic effect or as a result of the nontrivial
measure of the model.

We further pointed out that this model, and therefore also
the two-dimensional quantum Hall system, is described holographically
in terms of a one-dimensional system, the so-called Sutherland
integrable model of particles on the circle.

The correspondence between states and operators of the matrix 
and quantum Hall systems is still an open issue. In principle, such 
a correspondence is guaranteed to exist, since the two Hilbert
spaces have the same dimensionality, as demonstrated in the last
section. For it to be useful, however, it should be such that
operators in the quantum Hall system map into explicit, simple
matrix model operators. The hope is that the operators 
in the matrix model language will not involve explicitly the filling 
fraction (unlike, e.g., the hole creation operators in the 
second-quantized
fractional quantum Hall system), and thus will describe the
properties of these systems in a more universal way. This would
also open the road for the calculation of relevant quantities, 
such as correlation functions, in the matrix model formulation.
This is a most important issue.

There are clearly many other open questions. Incorporating the
spin of the electrons and identifying skyrmion-like configurations
is an obvious next step. Most intriguing, however, is the question
of a possible phase transition of the quantum Hall
system at small filling fractions. Numerical simulations suggest
that at $\nu^{-1} \sim 67$ Laughlin electrons form a Wigner crystal
instead of an incompressible fluid. This is based on the properties
of the Laughlin wavefunctions and does not seem to hinge on the
specific dynamics of the electrons beyond what is already encoded
in the
wavefunctions themselves. The corresponding one-dimensional
Sutherland system does not exhibit any such phase transition.
This does not guarantee, however, that two-dimensional quantities
calculated in the context of this model would not exhibit nonanalytic
(or at least crossover) behavior in the filling fraction, signaling
a phase transition. This intriguing possibility is the subject
of further research.

\vskip 0.2in
{\it Acknowledgements}: I would like to thank 
Dimitra Karabali and Bunji Sakita for illuminating discussions
on the correspondence between matrix model and Laughlin states,
Simeon Hellerman, Mark van Raamsdonk and Lenny Susskind for very
interesting discussions on the possible operator mapping 
between the models and the Wigner phase transition, and
Bogdan Morariu and Parameswaran Nair for useful  discussions
on the canonical and group theoretic structure of the model.
I would also like to thank for its hospitality the Physics Department
of Columbia University where part of this work was done.

\end{document}